\documentclass[useAMS,usenatbib]{mn2e}
\usepackage{graphicx}

\title[Optical Classification of M Dwarfs]{Optical Spectroscopic Classification and Membership of Young M Dwarfs in Star-Forming Regions}

\author[F. C. Riddick et al.]{F. C. Riddick$^{1,2}$, P. F. Roche$^{2}$, P. W. Lucas$^{3}$, \\
$^{1}$ Dept. of Astronomy \& Astrophysics, Penn State University, 525 Davey Lab, University Park, PA 16802, USA \\
$^{2}$ Astrophysics, University of Oxford, Dept. of Physics, DWB, Keble Road Oxford OX1 3RH, UK \\
$^{3}$ Centre for Astrophysics, University of Hertfordshire, College Lane, Hatfield, Herts, AL10 9AB, UK \\
}

\begin{document}



\maketitle

\begin{abstract}

The spectral type is a key parameter  in calibrating the temperature which is required to estimate the mass of young stars and brown dwarfs.  We describe an approach developed  to classify low-mass stars and brown dwarfs in the Trapezium Cluster using red optical spectra, which can be  applied to other star-forming regions. The classification uses two methods for greater accuracy: the use of narrowband spectral indices which rely on the variation of the strength of molecular lines with spectral type and a comparison with other previously classified young, low-mass objects in the Chamaeleon I star-forming region. We have investigated and compared many different molecular indices and have identified a small number of indices which work well for classifying M-type objects in nebular regions. The indices are calibrated for young, pre-main sequence objects whose spectra are affected by their lower surface gravities compared with those on the main sequence. Spectral types obtained are essentially independent of both reddening and nebular emission lines.

Confirmation of candidate young stars and brown dwarfs  as bona fide cluster members  may be accomplished with moderate resolution spectra in the optical region by an analysis of the strength of the gravity-sensitive Na doublet. It has been established that this feature is much weaker in these very young objects than in field dwarfs. A sodium spectral index is used to estimate the surface gravity and to demonstrate quantitatively the difference between young (1-2Myr) objects, and dwarf and giant field stars. 

\end{abstract}



\section{Introduction}
\label{sec:intro}
Mass is the most fundamental property of any astronomical object, governing its observed properties and lifetime's evolution. A common goal of many studies is to constrain the low-mass end of the initial mass function (IMF) and determine the possible mechanisms of brown dwarf (BD) formation. In the rare cases of eclipsing binaries, e.g. Stassun Matthieu \& Valenti (2006), or spatially resolved astrometric binaries, e.g. Bouy et al. (2004), accurate mass estimates may be obtained without reference to stellar models. In the majority of other cases mass must be obtained by less direct methods. In young clusters, where the members may be assumed to be approximately coeval, masses may be estimated by assigning an average age to all members and comparing the positions of the objects on the HRD with theoretical evolutionary tracks. 

Since young, low-mass objects are located on primarily vertical tracks on the HRD (e.g. Baraffe et al. 1998), mass estimates are much more sensitive to errors in effective temperature (T$_{\it eff}$) than in luminosity. Spectroscopy provides a better estimate of  $T_{\it eff}$ than photometric data can yield and hence more accurate estimates of mass may be derived. The optical region is ideally suited to spectral classification of M dwarfs since it contains many highly temperature sensitive molecular absorption bands which may be used for accurate classification. The rapid change in optical spectroscopic features with temperature enables \textit{relative} classifications to be very precise and accurate spectral types can be achieved from low-resolution spectra for faint M-type sources, e.g. Luhman (1999).  In contrast, \textit{absolute} classifications require careful consideration of the gravity-sensitivity of the features examined, as discussed in Section~\ref{sec:molecular-indices}.

Effective temperatures may be estimated from spectra in two ways: by comparison of the spectra with model spectra of known temperature (e.g. in the IR by Lucas et al. 2001; Natta et al. 2002) or by using an appropriate temperature scale for the transformation of spectral type to $T_{\it eff}$. We use the less direct latter method to obtain the best estimate of $T_{\it eff}$, therefore spectral classification needs to be as accurate as possible in order to obtain the most accurate mass estimates. This is achieved by using two complementary methods as described in Section~\ref{sec:class}. This method is preferable since classification by a direct comparison of observed and model spectra relies heavily on the model spectra and any inadequacies present in their construction will make classification unreliable. In particular, models at young ages suffer the most uncertainties (e.g. Baraffe et al 2003, Marley et al 2007). 

For statistical studies of a mass function or to obtain information about a cluster's star-formation history, it is essential to establish that all of the objects observed are actually cluster members and not foreground or background objects. Young BDs in star-forming regions (SFRs) like the Trapezium Cluster can have the same $T_{\it eff}$ or apparent magnitude as older, higher mass, field stars and photometric studies can provide no distinguishing evidence of cluster membership. Spectroscopy however can yield surface gravity indicators, since many spectral features are highly sensitive to surface gravity. Young PMS objects have surface gravities much lower than those of older field objects,  so that low gravity signatures give strong evidence of youth and hence confirmation of cluster membership.

We have investigated spectral indices to measure surface gravity semi-quantitatively in order to establish evidence of youth and hence cluster membership. These methods of spectral classification and membership analysis were designed to be applied to data for $\sim$ 40 late M type low-mass stars (LMS) and BD candidates in the Trapezium Cluster in Orion, which are described in a companion paper (Riddick, Roche \& Lucas submitted; hereafter Paper 2).

\subsection{Aims Of This Work}
We first devise a robust system to classify moderate resolution (R$> 1000$) red optical spectra from 6000-9000\,\AA\, of LMS and BDs in SFRs (Section~\ref{sec:class}). We discuss the possible methods of classification and their uses and limitations. Our method combines classification using several temperature-sensitive molecular spectral indices and a comparison with other young sources (Section~\ref{sec:class}). Secondly, we investigate spectroscopic evidence of cluster membership using surface gravity indicators (Section~\ref{sec:membership}).  

\section{Observations in Star-Forming Regions}
\label{sec:obs}
\subsection{Nebular Emission Lines}
A particular complication in star-forming regions is the bright nebulosity which may pervade young clusters.  The nebular spectrum still often appears in the extracted spectra of many objects -- in some to a much greater extent than others, due to the high spatial variability of the nebula. For example, nebular $H\alpha$ emission is present and strong in all of the Orion spectra discussed in Paper 2, making an assessment of its equivalent width as a membership criterion impossible. Our classification system was therefore designed to be unaffected by the presence of any remaining nebular emission lines in the spectra. 

\subsection{Reddening}
Dust reddens the spectra of many young objects, although this can be partially circumvented by selection of objects with  low extinction.  Nevertheless, to ensure the most accurate spectral typing possible for any object, the spectra must be dereddened since insufficient dereddening will make the overall slope of BDs and LMS spectra appear higher than it really is, possibly leading to later derived spectral types.  The procedure used for spectral typing (Section~\ref{sec:class}) was designed specifically to ensure that the results are as free as possible of the effects of imperfect dereddening on the spectra.  

\section{Optical Spectral Classification of M Dwarfs}
\label{sec:class}
\subsection{Spectral Features}
The optical spectra of M dwarfs are dominated by molecular absorption bands from metal oxide species such as titanium oxide (TiO) and vanadium oxide (VO). These have large opacity cross-sections per molecule, so dominate the spectra despite being trace species. The extreme temperature sensitivity of these metal oxide bands can be utilised for spectral classification and they are the primary temperature indicators in cool stars.
 
The TiO and VO bands increase in strength with decreasing temperature along the majority of the M sequence, both showing a smooth variation with spectral type, as shown by the spectral indices based on their strength in Figure~\ref{indices-plots} and Section~\ref{sec:molecular-indices}.
TiO is prominent for all M dwarfs whereas VO begins to dominate for M5 and later types and is a good temperature indicator for the latest objects (Kirkpatrick et al. 1993).  Both TiO and VO reach a maximum at very late M types and weaken rapidly beyond M types into the L sequence (Kirkpatrick et al. 1999). As $T_{\it eff}$ decreases below $\sim$ 2400K (Chabrier et al. 2000), TiO and VO lose their prominence due to dust grain condensation. Other elements with lower condensation temperatures such as the alkali metals and the hydrides FeH and CrH are relatively unaffected, favouring their detection (Allard et al. 2001).  

\subsection{Veiling}
For many objects, continuum veiling is expected to be fairly weak since accretion luminosity is generally a small fraction of stellar luminosity (Robberto et al. 2004) and the positions of the stars in the HRD remain almost unaffected by the accretion contribution.  However, veiling of narrow absorption lines can still be high, if emission lines that veil the stellar absorption lines are produced in a stellar wind. This will  be less important for most molecular lines than for atomic lines, though some species may be affected by circumstellar emission.  At infrared wavelengths, thermal emission from circumstellar dust can dilute molecular absorption bands.  This type of veiling can certainly affect spectral classification and must be taken into account since it can make a later type spectrum appear  earlier  by decreasing the depth of the molecular absorption lines used for classification, but the effect is likely to be small in the red spectral region considered here. 

\subsection{Spectral Typing by Comparison with Other Classified Spectra}

Very young objects, such as those in SFRs, which are still contracting to the MS have larger radii and therefore lower surface gravities than those of older field dwarfs of the same mass. Due to the gravity sensitivity of many features, the spectra of old field M dwarfs are different from much younger objects. 

It is possible to classify very young objects by comparison with field objects: those with spectral types earlier than M5 are best compared with dwarf spectra and those later than M5 with averages of dwarf and giant spectra (e.g. Luhman 1999). However the use of previously classified sources of a very similar age, rather than field dwarfs and giants, is preferable since the surface gravity changes considerably between the ages of even 1 and 10 Myr according to models (e.g. Burrows et al. 1997),  while giants later than M5 are spectrum variables and are thus unsuitable as spectral standards (Kirkpatrick et al. 1991). In addition, optically (rather than IR) classified young objects are preferred as standards, since the high temperature sensitivity of the molecular bands in the optical region provides the most accurate calibration of spectral types. The optically calibrated young objects can then be used as calibrator sources at infrared wavelengths.  In this work we use the classifications obtained by Luhman and co-workers (e.g. Luhman 2004, Luhman et al 2003, Briceno et al 2002) as the basis for our spectral types.

\begin{figure}
\includegraphics[height=4in]{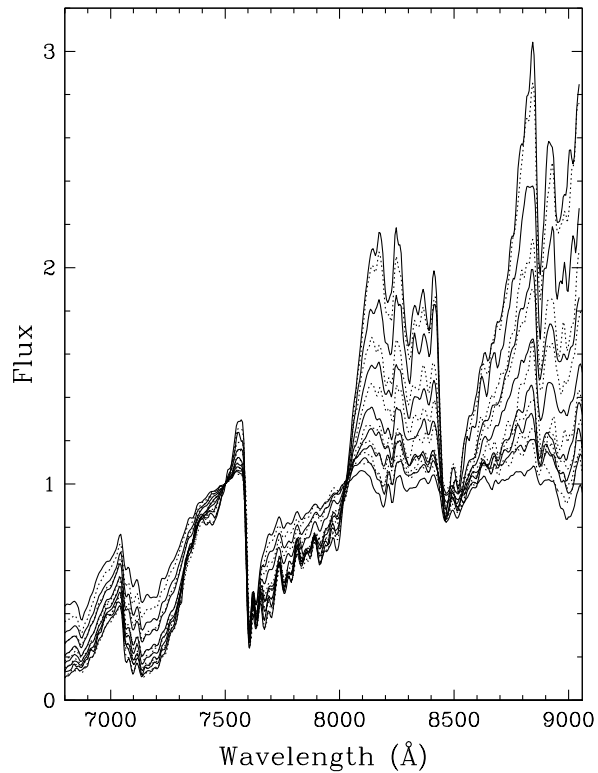}

\caption{Chamaeleon spectra from M3-M8 from Luhman (2004). Types shown are, from bottom to top: M3, M3.25, M3.5, M4, M4.5, M4.75, M5, M5.25, M5.5, M5.75, M6, M6.5, M7.25, M7.75 and M8. The spectra are normalised to 7500\,\AA. In addition to this point, the spectra also all coincide at $\sim$8025\,\AA\, and $\sim$8465\,\AA. This fact enables indices to be constructed which measure the rapidly and monotonically varying flux in this region with respect to these 2 fixed points. }
\label{cham-set}
\end{figure}

\subsection{Spectral Typing Using Indices}
\label{sec:indices-typing}
Spectra may also be classified using spectral indices which are calculated by taking ratios of the average flux within narrow wavelength regions. The indices are calibrated by reference to previously classified objects. The main benefit of using indices for classification is that precise types may be obtained without the need for a visual comparison of the spectra with others, since it is often difficult to distinguish spectra even half a subclass apart by eye, especially for low signal-to-noise (S/N) spectra. In addition, the indices may be calibrated for dwarfs, giants or PMS objects. A visual comparison can still be used in addition to this method to ensure accurate results.

Ideally, the relationship between an index and spectral type should be monotonic and discriminate over a large range of spectral subclasses.  In practice, many indices based on molecular features are not monotonic over large ranges of spectral types but instead peak at late-M types and decline thereafter as the bands which they measure first increase in strength with decreasing temperature due to molecule formation and then decrease due to dust formation and depletion from the gas phase.  If the indices are double-valued then their usefulness for spectral types near the turnover is limited. If different indices changed sign at different types, then it would still be possible to deduce the correct type using multiple indices, but many of the indices \textit{do} reverse at the same type, around M8-M9, at the onset of dust formation, as can be seen in Figure~\ref{indices-plots}. However, since the spectra are also compared with other previously classified sources before a final spectral type is assigned, any mis-typing can be avoided.  If objects to be classified span the M/L boundary then separate two-part fits to the indices are necessary.  

Molecular indices also have an upper $T_{\it eff}$ limit for spectral typing purposes due to the fact that the features measured appear in the spectra only at and below their temperature of formation.  Most of the indices we calculated flattened considerably at types earlier than M3 (See Tables~\ref{tab:fits} \& \ref{tab:fits-not-used} and Figure~\ref{indices-plots}), so this classification system will not produce accurate spectral types for these early M types. However, if the spectra are examined by eye, it is obvious if the type is earlier than this and so spectral typing can instead be accomplished by comparison with other classified sources.  

In J- and K-band spectra of substellar objects, the oxides most frequently used for M star classification (H$_{2}$O and CO) are relatively independent of surface gravity (Gorlova et al. 2003). In addition, the effects of surface gravity on spectral indices are less pronounced for M than L dwarfs (Barrado y Navascues et al. 2001). The indices we calculated in the optical region suggest that neglecting surface gravity effects on band strength would introduce spectral typing errors of up to 1 subtype (see Figure~\ref{indices-plots}), therefore for young objects indices which have been calibrated using other young sources rather than field dwarfs or giants should be used, in order to separate completely the effects of gravity and temperature on the spectral features and obtain the best possible estimates of  spectral type. 

\subsubsection{Molecular \& Atomic Indices} 
\label{sec:molecular-atomic}
The high temperature sensitivity of the molecular features makes use of indices based on the strength of these an obvious choice for classification and a large number have been defined and used for this purpose. Molecular indices measure ratios of the strengths of temperature sensitive molecular absorption features with respect to a region of nearby pseudocontinuum (due to the large number of overlapping molecular bands in M dwarf spectra, there is no true continuum remaining in the spectra). The depth of individual features with respect to the pseudocontinuum will not be affected significantly by reddening and this insensitivity to reddening is a major advantage of molecular indices.

Many atomic lines, such as those of the alkali metals, are also temperature sensitive and may be used for classification (e.g. Kirkpatrick et al. 1991). However, these lines are also highly sensitive to surface gravity, much more than to temperature, and are therefore unsuitable for classification purposes unless the surface gravity of an object is known. This may be the case for old field objects but not for very young objects in SFRs where the surface gravity is highly dependent on age -- see Section~\ref{table-notes}. 

\subsubsection{Pseudocontinuum (PC) Indices} 
These measure the PC slope in optical spectra by measuring the relative flux in different sections of PC, in effect a colour ratio, using the few available points of PC between the molecular opacity.  However, these indices are affected by reddening and  errors  in the dereddening procedure will lead to inaccurate spectral types.  This will be worse where the wavelength regions used for the indices are widely separated, as is the case for many of these indices, e.g. the PC5 index of Mart\'{i}n et al. (1996) in Table~\ref{indices-list}.  PC indices are therefore better suited to classification of low-mass field stars and BDs which are nearby and hence suffer minimal extinction. On the other hand, an advantage of PC indices is that they are relatively insensitive to variations due to features being affected by gravity, and they can be useful for data with low signal-to-noise ratio.  

\subsection{Indices for Optical Spectral Classification}
\label{sec:molecular-indices}
Several indices have been defined and used in previous studies in the wavelength range  
6,000--9,500\,\AA\, used here.  We have also defined and calculated some new indices calculated from the region around 8000-8465\,\AA. This region was chosen as there is prominent spectral structure which  shows a smooth variation with spectral type, as shown in Figure~\ref{cham-set} for a sample of Cha I objects, and it is possible to obtain high quality spectra because of high instrument sensitivity and good atmospheric transmission.  We have defined 4 different indices in this region to measure the depth of TiO absorption with respect to the pseudocontinuum.

\begin{figure}
\includegraphics[height=4in]{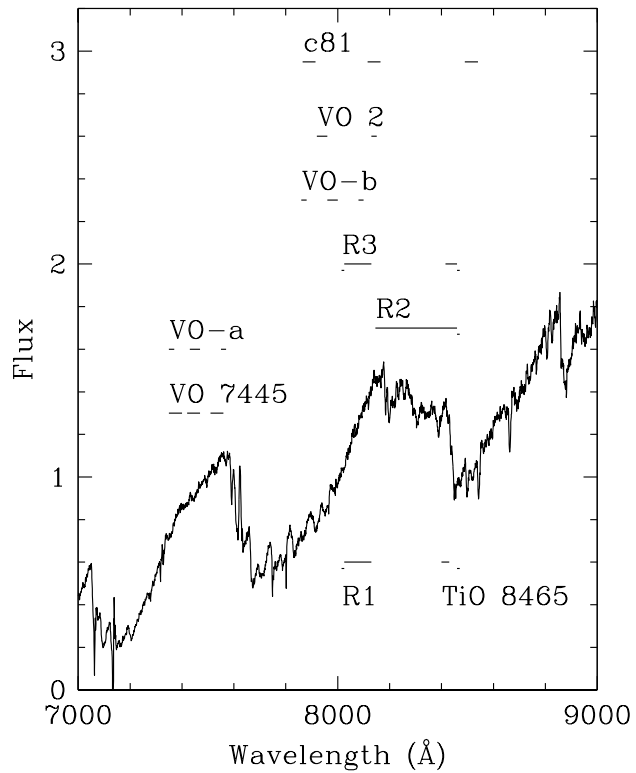}

\caption{Spectral Regions used in the 9 Spectral Indices used for spectral typing shown superimposed on the spectrum of an M5.25 object in the Trapezium Cluster}
\label{indices-regions}
\end{figure}

\begin{center}
\begin{table*}
\caption{\textbf{Optical Molecular and Colour Indices for Spectral Classification} See Section~\ref{table-notes} for notes.}
\label{indices-list}
\begin{tabular*}{460pt}{lllcc}
\hline
Index & Numerator (\AA) & Denominator (\AA) & Reference & Notes \\
\hline

R1         & 8025-8130	  & 8015-8025 & g & \\
R2         & 8145-8460	  & 8460-8470 & g & \\
R3         & (8025-8130)+(8415-8460) & (8015-8025)+(8460-8470) & g & \\
R4         & 8854-8857	  & (8454-8458)+(8873-8878) & g & \\
c81        & 8115-8165	  & (7865-7915)+(8490-8540) & i & \\
TiO 1     & 6718-6723  & 6703-6708	 & f & B \\
TiO 2     & 7058-7061  & 7043-7046	& f &B \\
TiO 3     & 7092-7097  & 7079-7084	& f &B \\
TiO 4     & 7130-7135  & 7115-7120	& f &B \\
TiO 5     & 7126-7135  & 7042-7046	& f &A \\
TiO 6     & 7550-7570	  & 7745-7765 & d & A \\
TiO 7       & 8440-8470 & 8400-8420 & a \\
TiO-a     & 7033-7048	  & 7058-7073 & c & B \\
TiO-b     & 8400-8415  & 8435-8470 & c & \\
TiO 8465 & 8405-8425 & 8455-8475 & h & \\
VO 1       & 7430-7470 & 7550-7570 &  a \\
VO 2      & 7920-7960  & 8130-8150 & d & F\\
VO 7445 & 0.5625 (7350-7400) + 0.4375 (7510-7560) & 7420-7470 & b & \\
VO-a      & (7350-7370)+(7550-7570) & 7430-7470 & c & \\
VO-b      & (7860-7880)+(8080-8100) & 7960-8000 & c & \\
CrH-a     & 8580-8600  &8621-8641 & c & \\
CrH-b     & 9940-9960  &9970-9990	& c & C \\
FeH-a     & 8660-8680  & 8700-8720 & c & \\
FeH-b     & 9863-9883  &9908-9928	 & c &C \\
CaH 2    & 6814-6846  & 7042-7046 & f &B \\
CaH 3    & 6960-6990  & 7042-7046 & f & B \\
Na-a       & 8153.3-8163.3 & 8178.3-8188.3	& c &D \\
Na-b       & 8153.3-8183.3 & 8189.8-8199.8	& c & D \\
Rb-a       & (7775.2-7785.2) + (7815.2-7825.2) & 7795.2-7805.2 & c & D \\
Rb-b       & (7922.6-7932.6) + (7962.6-7972.6) & 7942.6-7952.6 & c &D \\
Cs-a   	   & (8496.1-8506.1) + (8536.1-8546.1) & 8516.1-8526.1 & c &D \\
Cs-b        & (8918.5-8928.5) + (8958.3-8968.3) & 8938.5-8948.5 & c &D \\
PC1        & 7030-7050  & 6525-6550	 & e & B \\
PC2        & 7540-7580  & 7030-7050	 & e & \\
PC3        & 8235-8265  & 7540-7580	 & e & \\
PC4        & 9190-9225  & 7540-7580 & e & C,E \\
PC5        & 9800-9880  & 7540-7580	 & e &C,E \\
colour-a  & 9800-9850  &7300-7350 & c & C,E \\
colour-b  & 9800-9850  &7000-7050 & c & C,E \\
colour-c  & 9800-9850  &8100-8150 & c & C,E \\
colour-d  & 9675-9875  &7350-7550	& c & C,E \\
\hline

References:\\
a Hawley et al. (2002)\\
b Kirkpatrick et al. (1995)\\
c Kirkpatrick et al. (1999)\\
d Lepine et al. (2003)\\
e Mart\'{i}n et al. (1996)\\
f Reid et al. (1995)\\
g Riddick (2006)\\
h Slesnick et al. (2006) \\
i Stauffer et al. (1999)\\
\end{tabular*}
\end{table*}
\end{center}

\subsubsection{Notes on Indices in Table~\ref{indices-list}}
\label{table-notes}
The indices listed in Table~\ref{indices-list} may be affected by several factors which may make them produce inaccurate spectral types: 

A. Some of these indices can be rendered useless (e.g. in Orion) by the inclusion in their wavelength regions of strong nebular emission lines, since the index would then measure the strength of the nebular line rather than the feature defined.  Many of the Trapezium spectra contained no residual nebular features, but a significant number did and so for the sake of consistency these indices were not used for spectral typing.  Indices coinciding with nebular emission lines are TiO 5 and TiO 6. These indices may still be of use in less nebulous regions.

B. Wavelength regions below $\sim$ 7200\,\AA.  Since M type spectra have relatively little flux here, indices measuring relative flux at short wavelengths are subject to larger random errors. This applies to: PC1, TiO 1, TiO 2, TiO 3, TiO 4, TiO-a, CaH 2 and CaH 3. TiO 5 is also at a short wavelength but gives a reasonable correlation with spectral type.

C. Wavelength regions above $\sim$ 9000\,\AA. Our spectra were affected by low S/N due to the rapidly declining CCD response, making indices in this region less suitable for classification: CrH-b, FeH-b, PC5 and the 4 colour indices.

D. Gravity Sensitivity.  The alkali metal lines are much more sensitive to surface gravity than temperature and are therefore not suitable for spectral typing of very young objects. This applies to the Na, Rb and Cs indices.

E. Sensitivity to Reddening. The indices which have the longest baselines are most affected by reddening, e.g. PC4, PC5 and the colour indices.

F. We note that this is very similar to an index defined by Wilking et al. (2005)-- similar regions are measured, but different width bands are used.

\begin{center}
\begin{table}
\caption{\textbf{Chamaeleon Template Objects :  Spectral Types and Indices} }
\label{templates}
\begin{tabular*}{8.2cm} {llllll}

\hline
ID & SpT & VO7445 & c81 & VO 2 & Na C \\
\hline
T56 & M0.5 & 0.995 & 1.020 & 0.964 & 1.101\\
CHXR 54 & M1& 0.982 & 0.915 & 1.030 & 1.118\\
CHXR 40 & M1.25 & 0.991 & 1.025 & 0.971 & 1.098\\
T20 & M1.5 & 0.996 & 1.059 & 0.932 & 1.126\\
CHXR 14s & M1.75 & 0.987 & 1.041 & 0.969 & 1.145 \\
T39 & M2 & 0.994 & 1.105 & 0.896 & 1.101\\
CHXR 68B & M2.25 & 0.989 & 1.036 & 0.953 & 1.148 \\
CHXR 48 & M2.5 & 0.992 & 1.082 & 0.912 & 1.127\\
B53 & M2.75 & 0.988 & 1.046 & 0.963 & 1.139\\
CHXR 71 & M3 & 0.985 & 1.100 & 0.906 & 1.146\\
T05 & M3.25 & 0.984 & 1.163 & 0.842 & 1.085\\
Hn 18 & M3.5 & 0.985 & 1.186 & 0.815 & 1.135\\
T10& M 3.75 & 0.981 & 1.214 & 0.776 & 1.123\\
Hn 17 & M4 & 0.982 & 1.225 & 0.784 & 1.137\\
T12 & M4.5 & 0.988 & 1.343 & 0.719 & 1.056 \\
T55 & M4.5 & 0.989 & 1.392 & 0.693 & 1.081\\
CHXR 35 & M4.75 & 0.993 & 1.457 & 0.650 & 1.098\\
Hn 07 & M4.75 & 0.983 & 1.439 & 0.663 & 1.092\\
Hn 02 & M5 & 0.977 & 1.479 & 0.640 & 1.056\\
T50 & M5 & 0.987 & 1.565 & 0.612 & 1.025\\
CHXR 15 & M5.25 & 0.986 & 1.541 & 0.606 & 1.037 \\
T37 & M5.25 & 0.989 & 1.597 & 0.583 & 1.049\\
CHXR 84 & M5.5 & 1.004 & 1.611 & 0.569 & 1.065\\
Hn 12W & M5.5 & 1.000 & 1.619 & 0.576 & 1.048\\
Hn 13 & M5.75 & 1.010 & 1.763 & 0.502 & 1.001\\
Cha H$\alpha$ 6 & M5.75 & 1.022 & 1.751 & 0.511 & 1.023\\
CHSM 1982 & M6 & 1.041 & 1.862 & 0.471 & 1.005\\
Cha H$\alpha$ 10 & M6.25 & 1.050 & 1.906 & 0.456 & 1.006\\
ISO 138 & M6.5 & 1.061 & 2.007 & 0.439 & 1.019\\
Cha H$\alpha$ 11 & M7.25 & 1.103 & 2.094 & 0.398 & 0.967\\
Cha H$\alpha$ 1 & M7.75 & 1.107 & 2.233 & 0.351 & 0.945\\
CHSM 17173 & M8 & 1.140 & 2.353 & 0.334 & 0.922\\
 
\hline
\end{tabular*}
\end{table}
\end{center}

\begin{figure*}


\includegraphics[width=6in]{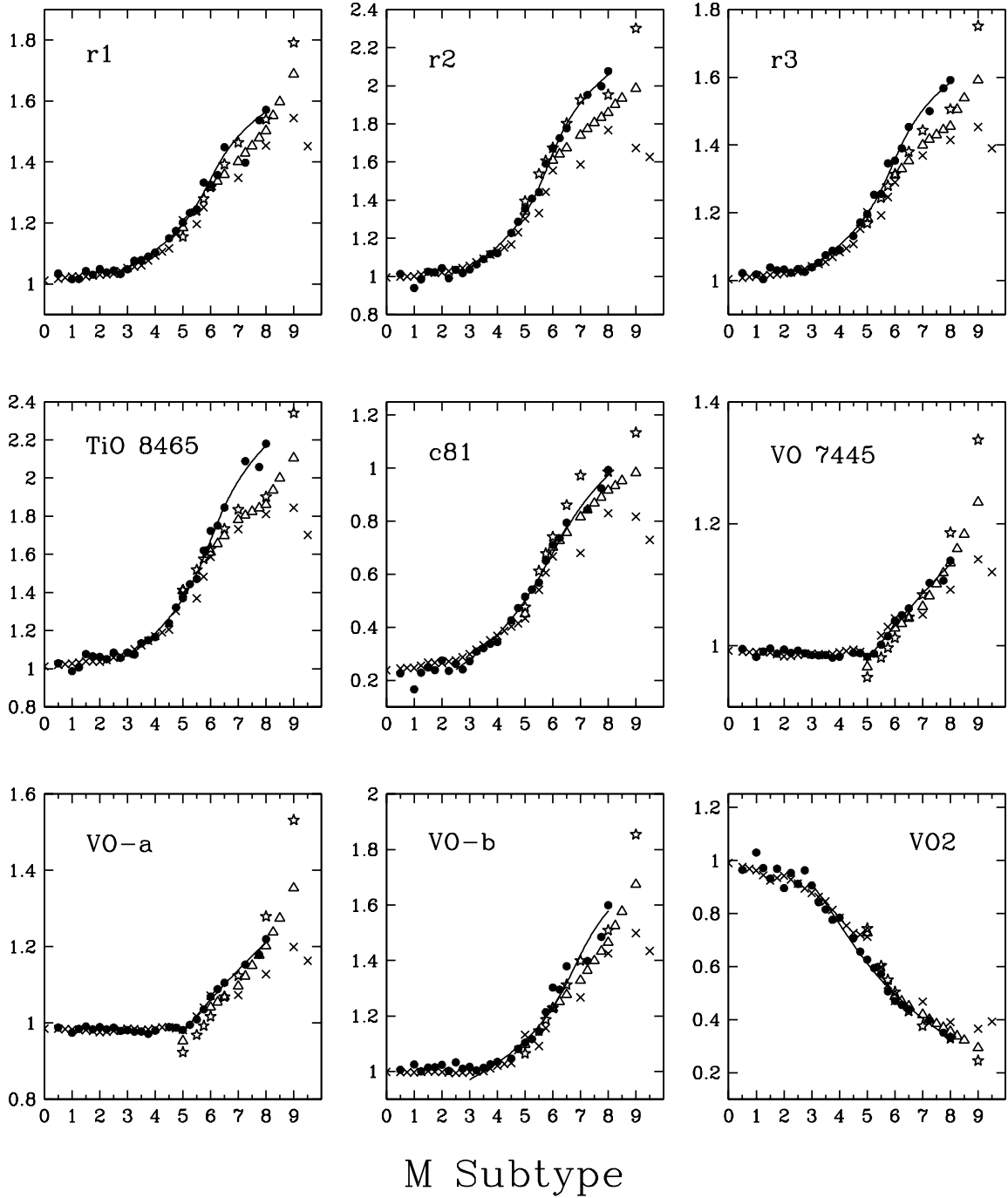}
\caption{\textbf{Fits for Indices Used for Spectral Typing.} Indices are plotted for Chamaeleon I objects (filled circles), field dwarfs (crosses), giants (stars) and dwarf/giant averages (triangles). The line of best fit to the Chamaeleon I points is used to calibrate the indices for classification of young PMS objects in SFRs.  The line is extrapolated to early and late spectral types as a dotted line to illustrate the spectral types at which the index is no longer valid for classification: early types are not necessarily valid due to the invariability of the indices and later spectral types due to the Chamaeleon objects used to obtain the relationship only going as late as M8.}
\label{indices-plots}

\end{figure*}

\subsubsection{Spectral Typing Procedure using Indices}
\label{procedure}
In order to use indices for spectral typing, it is necessary to calibrate them against spectra of objects of known type. To provide an accurate calibration and average out the effects of small changes in the index strengths due to the discretisation of the spectral types of the standards, we used several objects of each spectral subtype where possible. 

We calculated the indices in Table~\ref{indices-list} for the spectra of 32 low-mass stars and BDs from M0.5--M8 in the 2 Myr old Chamaeleon I SFR, (Luhman 2004), 12 in the 1 Myr old Taurus SFR, from M5.25--M9.5 (Brice\~no et al. 2002; Luhman et al. 2003) and for a large sample of 48 field M dwarfs, giants and dwarf/giant average spectra from Kirkpatrick et al. (1991), Henry et al. (1994) and Kirkpatrick et al. (1997), as used by Luhman (1999), which together contain a very large sample of M and L type objects of all luminosity classes, the standard sub-types being represented by averages of 1 or more stars. The spectral types of the members of Chamaeleon I were derived by Luhman (2004) by comparisons to dwarfs and averages of dwarfs and giants for types earlier or later than M5, respectively (as in Luhman 1999). 

The value of the index was then plotted against spectral type. Many indices did not produce obvious correlations with spectral type for the reasons given in Section~\ref{table-notes} and these were used no further.  The remaining 19 indices did show a good correlation with spectral type (see Tables~\ref{tab:fits} \& \ref{tab:fits-not-used}). Polynomial relationships between the spectral type of the Chamaeleon objects and these indices were derived using DIPSO in order to calibrate the indices. The fits, whose orders varied between 2 and 4, were calculated only from spectral types later than those given in column 2, as discussed in Section~\ref{sec:indices-typing}, so the fit is only valid for  objects later than this type although note that  the fit is extrapolated and shown for earlier M types as a dotted line. 

The latest Chamaeleon object was an M8 object, therefore the relationship obtained from these was not calculated for types later than M8 and is not valid at later types, although this is also shown as a dotted line in Figure~\ref{indices-plots}. However an examination of the indices in Figure~\ref{indices-plots} shows that the dwarf/giant average  for the VO2 index agrees well with the Chamaeleon data for all M spectral types extending beyond M8,  and providing a continuation of the relationship up to M9.5.  It appears that this index can be used for all M-types later than M3.   In addition, very few types later than M8 were found in the Orion data and these all had low S/N, making spectral typing using indices less reliable.

Some of these indices showed distinct series for the field dwarfs, giants and dwarf/giant averages indicating a variation of the strength of the features with surface gravity. The indices of the Chamaeleon and Taurus objects lay in general at either intermediate values or much closer to the giant than dwarf values. The Taurus data showed more scatter and contained fewer points so these were not used for calibration of the indices. The distinct series for the Chamaeleon objects and field M dwarfs highlights the importance of using objects of a similar age for index calibration and demonstrates that field dwarfs are unsuitable for spectral typing by comparison for these very young objects. Since the expected average age of the Trapezium objects is similar to that of the Chamaeleon objects, these were used for calibration of the indices. 

The  Chamaeleon objects used as templates are listed  in Table~\ref{templates} with the names and spectral types from Luhman (2004) and  the measured VO~7445, VO~2 and c81 and  Na  indices.   Other properties of these objects, such as the H-alpha fluxes and equivalent widths may be found in Table 2 of Luhman (2004).

\begin{center}
\begin{table*}
\caption{\textbf{Fits for Indices for Spectral Typing} 'x' is the value of the index calculated from the spectra}
\label{tab:fits}
\begin{tabular*}{16.5cm} {lll}

\hline
Index & Range of Validity &	 M Spectral Type-Index Relation  (x = index value) \\
\hline

VO 7445 &	M5--M8 &	type = 5.0881 + 17.121(x-0.982) + 13.078(x-0.982)$^{2}$\\
VO-a &	M5--M8 &	type = 5.0705 + 11.226(x-0.982) + 6.7099(x-0.982)$^{2}$\\
VO-b &	M3--M8 &	type = 3.4875 + 29.469(x-1.017) - 156.53(x-1.017)$^{2}$ + 394.28(x-1.017)$^{3}$ - 325.44(x-1.017)$^{4}$ \\
VO 2 &	M3--M8 &	type = 2.6102 -7.9389(x-0.963) -8.3231(x-0.963)$^{2}$ -14.660(x-0.963)$^{3}$\\
R1 &	M2.5--M8 & 	type = 2.8078 + 21.085(x-1.044) - 53.025(x-1.044)$^{2}$ + 60.755(x-1.044)$^{3}$\\
R2 &	M3--M8 &	type = 2.9091 + 10.503(x-1.035) - 14.105(x-1.035)$^{2}$ + 8.5121(x-1.035)$^{3}$\\
R3 &	M2.5--M8 &	 type = 2.8379 + 19.708(x-1.035) - 47.679(x-1.035)$^{2}$ + 52.531(x-1.035)$^{3}$\\
TiO 8465 &	M3--M8 &	type = 3.2147 + 8.7311(x-1.085) - 10.142(x-1.085)$^{2}$ + 5.6765(x-1.085)$^{3}$\\
c81 &	M2.5--M8 &	 type = 2.4331 + 8.0558(x-1.036) - 6.8171(x-1.036)$^{2}$ + 3.0567(x-1.036)$^{3}$\\
 
\hline
\end{tabular*}
\end{table*}
\end{center}

\begin{center}
\begin{table*}
\caption{\textbf{Fits for Additional Indices} These indices were not used for spectral typing since they produced less consistent types than those in Table~\ref{tab:fits-not-used}.}
\label{tab:fits-not-used}
\begin{tabular*}{16.5cm} {lll}

\hline
Index & Range of Validity &	 M Spectral Type-Index Relation  (x = index value) \\
\hline

PC2 &	M2--M8 &	type  =  0.95891 + 8.7890(x - 1.156) - 6.0903(x - 1.156)$^{2}$ + 1.8529(x - 1.156)$^{3}$\\
PC3 &	M3--M8 &	type = 2.0395 + 24.61(x - 0.956) - 50.292(x - 0.956)$^{2}$ + 39.489(x - 0.956)$^{3}$\\
TiO 5 &	M3--M8 &	type = 0.47587 - 14.907(x-0.862) - 28.702(x-0.862)$^{2}$ - 37.843(x-0.862)$^{3}$\\
CrH-a & M3--M8 &	type = 1.4801 - 46.29(x-1) - 37.824(x-1)$^{2}$\\
TiO-a &	M3--M8 &	type = 1.0810 + 9.8651(x-1.05) - 4.9939(x-1.05)$^{2}$ + 1.4823(x-1.05)$^{3}$\\
TiO-b &	M3--M8 &	type = 2.9069 + 13.841(x-1.067) - 23.646(x-1.067)$^{2}$ + 17.787(x-1.067)$^{3}$\\
R4 &	M3--M8 &	type = 2.7907 + 11.124(x-1.065) - 14.937(x-1.065)$^{2}$ + 10.609(x-1.065)$^{3}$\\
TiO 6 &	M0--M8 &	type = 0.91754 + 14.578(x-1.108) -27.808(x-1.108)$^{2}$ + 26.199(x-1.108)$^{3}$ -7.9521(x-1.108)$^{4}$\\
TiO 7 &	M2--M8 &	type = 2.2993 -27.642(x-0.941) -94.586(x-0.941)$^{2}$ -140.74(x-0.941)$^{3}$\\
VO 1 &	M0--M8 &	type = 1.0285 -52.770(x-0.961) -205.48(x-0.961)$^{2}$ -349.44(x-0.961)$^{3}$\\

\hline
\end{tabular*}
\end{table*}
\end{center}

\subsubsection{Application of indices to Orion data}
\label{Oriondata}

The 19 indices listed in Tables 3 and 4 were calculated for the Trapezium spectra and the relationships derived from the Chamaeleon data used to calculate the spectral type. However, many indices gave results inconsistent with the majority of the other indices (due to the reasons given in the notes to Table~\ref{indices-list}) and were therefore  not used for determination  of the spectral type. 

Nine indices remained which gave generally very consistent types and these were used for classification. See Table~\ref{tab:fits}. Fits for those indices giving less consistent types are given in Table~\ref{tab:fits-not-used}. The PC3 index of Mart\'{i}n (1996) gave a smooth variation with spectral type but no PC indices were used for the classification of the Trapezium objects, to avoid the uncertainties associated with dereddening, and instead molecular indices were used exclusively. 

In some cases the types were obtained with averages of less than these nine indices, due to defects in the data due to bad columns, chip gaps etc. which would affect the type obtained, but for the majority all nine indices were used. Spectral types were calculated from the mean of the remaining indices' types and compared.  Agreement between the types obtained from different indices was generally very good, with variations generally less than 0.25 subclasses.  The results from individual indices were examined to investigate if any produced consistently earlier or later types.  Only very small biases were found, the largest of these being for V0-b which produced a type on average 0.6 subtypes earlier than the average type.  The other indices had biases generally much smaller than 0.5 subtypes and in both directions.  

The success of typing from spectral indices varied greatly with the S/N of the spectra.  Very noisy spectra generally gave types from each index which differed considerably from each other and so were not assumed to be reliable (due to the low S/N and small wavelength regions used). These spectra were also usually untypable by eye. 

The nine indices cover three spectral regions with different wavelength ranges and bandwidths.  Two indices,  VO-a and VO 7445, cover the 7445\,\AA\,  VO band and have very similar dependences upon spectral type.  VO 7445  employs broader bands, and so is to be preferred.    The other 7 indices cover the  prominent structure between 7800 and 8500~\,\AA, either treating the edges at $\sim$8050 and 8400\,\AA\, separately or combining them with a measure of the pseudocontinuum.  Many of these again have similar dependences, with TiO~8465 showing the largest dependence on surface gravity and VO2 the smallest.   A combination of the VO 7445, VO2 and C81 indices will give good estimates of spectral type, although VO 7445 (and VO-a) is only sensitive at types later than M5, while TiO 8465 may provide some indication of gravity for very late spectral types.

\subsection{Final Spectral Types}
Each Trapezium object spectrum was also compared visually with spectra of all types from the Chamaeleon data (and Taurus for \textgreater M8). The spectra were normalised to 7500\,\AA\, and the overall shapes of the spectra were compared, in particular the region around 8000-8500\,\AA\, and the VO absorption at 7445\,\AA.

An independent assignation of spectral type was made based on these comparisons alone and this was then compared with the type obtained from the indices.  In the majority of cases, agreement between these 2 methods was within 1 subtype, giving weight to the validity of the index method of classification.  When the types differed, final types were given on a case by case judgement.  For example, in some cases, where spatial structure in the background left  significant residual flux  levels, this resulted in the types obtained from the indices being too early, therefore the types were taken from the visual assessment.  In the majority of cases, however, the difference is merely due to random errors introduced due to the difficulty of distinguishing such similar spectra.  

The classification system works well, giving smooth variations with spectral type and hence accurate types. The overall error in spectral typing is dominated by the uncertainty of the surface gravity of the object: is it really intermediate between that of dwarfs and giants, or is it more dwarflike or giantlike? We  assume that the young Chamaeleon sources are the most similar, having been classified by comparison with dwarfs and averages of dwarfs and giants for types earlier and later than M5 respectively. Estimated errors were 0.5 subtypes for types earlier than M7 and $\pm 1-2$ subtypes for later types due to these spectra having lower S/N.  These results are discussed fully in Paper 2.

The indices recommended for classification of low mass stars and BDs, and particularly the VO2 and VO 7445 indices, measure molecular absorption bands within relatively narrow and adjacent spectral bands to minimise the impact of reddening uncertainties. However,  it is important to consider the effects of errors in reddening corrections.  To investigate the possible errors incurred by imperfect dereddening, the spectral types obtained using spectral indices from unreddened  and reddened spectra were compared. On average, for spectra with A$_{V}$~= 10~mag, the spectral type obtained was $\sim$ 0.5 subtypes in error. The majority of the Trapezium objects have A$_{V}$ $\textless$ $5$~mag, and errors in dereddening significantly smaller than this and probably $\sim$1~mag so that the resulting errors in spectral typing due to imperfect dereddening will be small and  typically $<$~0.1 subtypes.

\section{Cluster Membership}
\label{sec:membership}
\subsection{Introduction}
For meaningful studies of cluster populations, determination of cluster membership is essential, but  establishing membership is difficult, even in relatively nearby clusters.   Foreground stars will generally have low extinction, but some genuine clusters members may do so too.   Background dwarfs have earlier spectral types for their fluxes, falling below the cluster temperature-magnitude sequence on an HRD, but these may not be easily distinguished from older cluster members if star formation has extended over several million years.   

\begin{figure}

\includegraphics[height=3in]{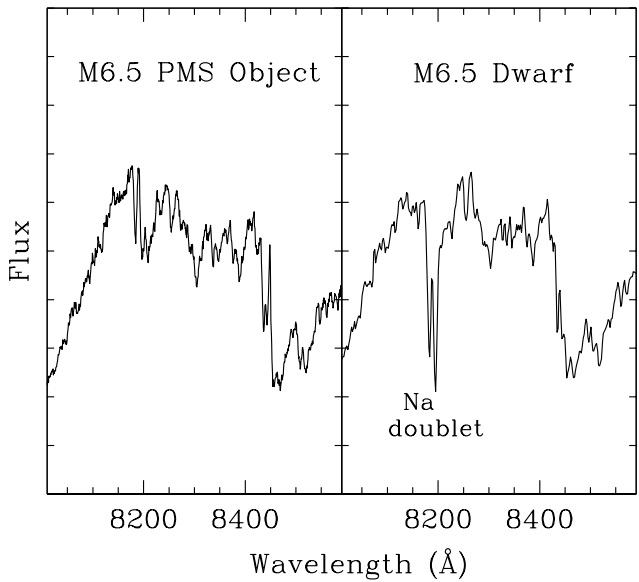}

\caption{Na Doublet absorption depth dependence on surface gravity illustrated in a young (left) and field (right) M6.5 star. }
\label{fig:na1}
\end{figure}

\subsection{Signatures of Youth}
Evidence of youth provides strong evidence of cluster membership.  Such evidence includes hydrogen recombination lines (e.g. H$\alpha$) which are strong in young sources, but also seen in older flare stars (e.g. Barrado y Navascues \& Martin 2003), and this is of very limited use in star-forming regions since HII region emission is likely to dominate the stellar emission lines.   Strong X-ray emission may also be present (Getman et al. 2005).

The presence of a circumstellar disk surrounding an object is also evidence of youth, disk lifetimes being  a few Myr (e.g. Haisch, Lada \& Lada 2001).  Disks may be evidenced in the optical region by strong Ca II emission at 8498, 8542, 8662\,\AA. This often arises in accretion onto a disk and is therefore evidence of youth and hence cluster membership. Other emission lines provide signatures of accretion or outflows and hence youth but many are likely to be contaminated by nebular emission lines in vigorous star-forming regions. 
 
The detection of Li I (e.g. at 6708\,\AA) can be used to establish cluster membership (e,g, Stauffer et al. 1999).  Lithium is destroyed in stars  more massive than 0.55 $M_{\sun}$ (Rebolo et al. 1992) but is expected to be present in very young BDs such as most of those in the Trapezium Cluster (\textless 10 Myr).   The region around Li I is contaminated by bright [SII]  $\lambda$6716/6731\,\AA\, nebular emission in Orion at moderate spectral resolution, and so Li absorption detection is therefore not useful as evidence for cluster membership in this case. 

Signatures of low gravity are the main form of evidence of youth (and hence cluster membership) readily available from these spectra and this is discussed in the following section.

\subsubsection{Gravity Sensitive Features}
For field dwarfs which have completed their contraction to the MS, the radius varies only slightly with age and so their high surface gravity depends simply on mass. For younger objects, the radius is several times larger than the eventual equilibrium value as they are still contracting to the MS, so they have lower surface gravities than MS objects of the same $T_{\it eff}$ (see e.g. Burrows et al. 1997). Young BDs can have the same $T_{\it eff}$ and apparent magnitude as old field dwarfs but if the gravity can be inferred from the spectrum this will break the degeneracy and enable young BDs, e.g. those in SFRs, to be distinguished from old field objects.  It is therefore extremely useful to be able to deduce the surface gravity of the object. Once spectral typing has been carried out using relatively gravity insensitive features such as the metal oxide bands, other more gravity sensitive spectral lines can be analysed to infer the surface gravity and provide evidence of cluster membership.

The alkali metal lines are highly gravity sensitive and useful for distinguishing PMS stars from field dwarfs (e.g. Martin et al 1999). The Na I doublet D lines at 8183/8195\,\AA\, are strong (due to the much higher abundance of Na than the other alkali metals) and are extremely gravity sensitive. M dwarfs, M giants and PMS M type objects have very different Na equivalent widths and absorption is much stronger with increasing surface gravity, i.e. dwarfs have the strongest absorption and giants and PMS objects have much weaker Na absorption, as can be seen in Figure~\ref{fig:na1}. Na strength considerably weaker than that found in dwarfs is therefore strong evidence of youth and hence cluster membership (e.g. Comeron et al. 1999). These differences are easily detected in low-resolution spectra,  and an accurate analysis of the lines is unnecessary for the purpose of distinguishing dwarfs from giants (e.g. Brice\~no et al. 2002). In this way cluster members may be easily distinguished from field dwarfs.

\begin{figure}
\includegraphics[height=4in]{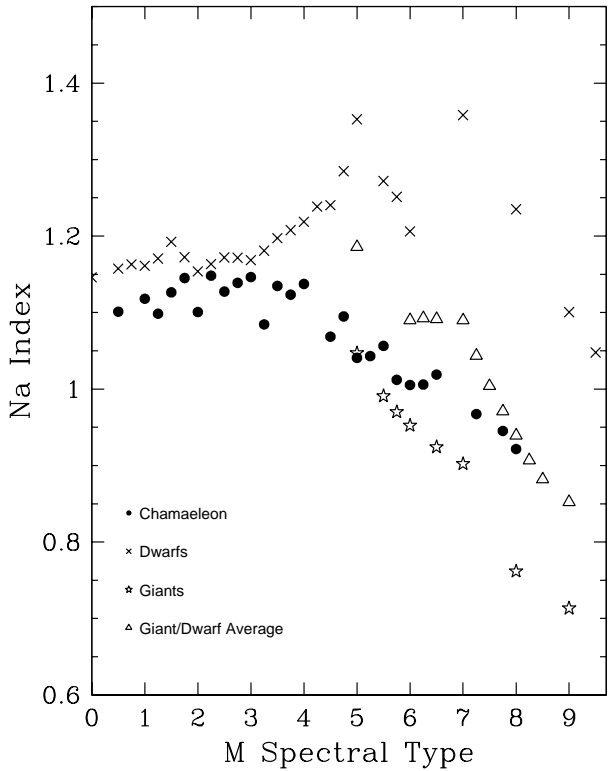}
\caption{Variation of Kirkpatrick et al. (1991) Na ratio with spectral type for Cha I objects and M-type dwarf, giant and combined standards }
\label{na-all}
\end{figure}

\subsubsection{Spectral Indices for Luminosity Classification}
Due to the large variation of Na doublet absorption with surface gravity, spectral indices based on its strength may be extracted and used to infer membership.  Kirkpatrick et al. (1991) defined 4 spectral indices, measuring the strength of CaH (6965), TiO (7358), Na I (8183,8195) and Ca II (8542) to classify spectra and measure gravity sensitive features and determine the luminosity class.  They found that all four indices showed a variation with M spectral type, many peaking at M5-7 types and then declining, but in each case there is a different sequence for dwarfs and giants, indicating that it is possible to distinguish between dwarfs, giants or PMS objects. 

%
%

We calculated these 4 indices for the standard stars used for the classification (Section~\ref{sec:molecular-indices}) and found no strong correlation of spectral type with the CaH, TiO and Ca II ratios.  This is to be expected since the CaH and TiO ratios are at a short wavelength where the low flux leads to greater uncertainties on the ratios, and ratio D the Ca II ratio is based on a feature which is highly dependent on accretion, much more than on spectral type, therefore it is not a reliable indicator for such young sources where accretion may be present.  

The Na ratio did show a trend with spectral type as seen in Figure~\ref{na-all}. The Na strength as measured by this ratio is very similar for Chamaeleon I objects and field M dwarfs for types earlier than M3, with the dwarfs showing marginally higher absorption. For types later than M3, the relative strength of Na differs significantly from dwarfs to giants and there is a marked decrease in Na absorption for the Cha I objects, with absorption between that of giants and giant/dwarf averages. At all M spectral types there is less absorption for the young objects, but only at types later than M3 is this unambiguous as evidence of youth and cluster membership. The ratio is therefore extremely useful for identifying young LMS and BDs in SFRs. Previous studies in SFRs have also found Na absorption to be significantly weaker in PMS objects than in field dwarfs, e.g. Gorlova et al. (2003) and Slesnick et al. (2004) find that atomic absorption lines in the optical and NIR are systematically weaker in young cluster BDs than in field stars.

\section{Summary}
We have defined a spectral classification system to classify young ($\sim$ 1 Myr) low mass stars and brown dwarfs with spectral types in the range M3-M9  in star-formation regions. This scheme provides spectral types, effectively independent of reddening and nebular line emission,   to an accuracy better than 1 subtype.  We recommend indices that sample the spectra near 7445, 8000, and 8440\,\AA\, using the VO 7445 index defined by Kirkpatrick et al (1995), the VO~2 index of Lepine et al (2003), the c81 index of Stauffer et al (1999). 

We have also analysed Na I doublet absorption strength quantitatively using a spectral index and found this to be a very good indicator of surface gravity, clearly separating the young objects from field dwarfs and giants and hence very useful for providing evidence of cluster membership.  The Na index clearly separates objects in the Chamaeleon clouds, where the ages are believed to be $\sim$2~Myr (Luhman 2004) from the field stars.  The TiO 8465 index defined by Slesnick et al (2006) also provides some gravity discrimination for spectral types later than M6. 

\section*{Acknowledgements}

We are very grateful to Kevin Luhman who generously provided the template spectra that underpinned this work.  FCR acknowledges the support of a PPARC studentship. 


\end{document}